\documentclass[aps,12pt,prb]{revtex4}

\usepackage{enumerate}

\usepackage{amsmath}

\begin{document}

\title{Impenetrable barriers and canonical quantization}

\author{Piotr Garbaczewski}

\author{Witold Karwowski}

\affiliation{Institute of Physics,
University of Zielona G\'{o}ra, PL-65 516 Zielona G\'{o}ra,
Poland}

\begin{abstract}
We address an apparent conflict between the traditional
canonical quantization framework of quantum theory and spatially
restricted quantum dynamics when the translation invariance of
an otherwise free quantum system is broken by boundary
conditions. By considering the example of a particle in an
infinite well, we analyze spectral problems for related confined
and global observables. In particular, we show how we can interpret
various operators related to trapped particles by not ignoring the rest
of the real line that is never occupied by a particle.
\end{abstract}


\maketitle

\section{Introduction}
A proliferation of papers on the pedagogical and more formal
aspects of the most idealized trapping model, the infinite
potential
well,\cite{stroud,hu,robinett,birula,valent,gori,klauder,gieres}
sophisticated exercises in exact quantization on a
half-line,\cite{voros} and the quantum mechanical approach to particles
on surfaces with obstacles,\cite{rembielinski} motivates renewed
interest in reconciling the principles of canonical quantization with
the analysis of well posed, spectral problems for the Hamilton operator
with Dirichlet boundary conditions.

The purely spectroscopic analysis is represented in the literature on
mesoscopic systems such as quantum billiards or microwave
cavities.\cite{stock,hurt,chavel} In this analysis one avoids using
canonical quantization and instead focuses on the statistical properties
of the related Laplace operator eigenvalues. Issues such as the position
and momentum observables and the indeterminacy relations are omitted
from the analysis of these spatially trapped quantum systems.

A major surprise in this context is that a careful analysis of the
conceptual background reveals unexpected inconsistencies and
paradoxes.\cite{valent,gori,klauder,gieres} They appear when one
applies the traditional apparatus of canonical quantization to
models of trapping and arise from attempts to give a correct
meaning to the differential expression $-i\hbar d/dx$. It is
possible to define different self-adjoint operators by means of
the same differential expression that leads to conflicting
options (compare Refs.~\onlinecite{valent,klauder,gieres} and
Refs.~\onlinecite{cohen,robinett,doncheski}) for what should be the
momentum observable and consequently the momentum representation
of wave functions for a particle in the infinite well.

The textbook canonical quantization procedure for a particle in
one spatial dimension is carried out in the Hilbert space $L^2(R)$
of square integrable functions on the real line $R$. The
canonical position and momentum operators $(Xf)(x)=xf(x)$,
$(Pg)(x)= -i\hbar {\frac{d}{dx}}g(x)$ are defined to act on
appropriate sets of functions $f,g \in L^2(R)$. If the motion of
the particle remains confined to a segment $[a,b]\subset R$, then
the corresponding wave functions are supported by $[a,b]$ and thus
form a subspace of $L^2(R)$. This subspace may be identified with
$L^2([a,b])$, the Hilbert space of square integrable functions on
$[a,b]$.

Therefore, for spatially confined dynamics, it appears
natural to neglect the (irrelevant) complement $R\setminus(a,b)$
of the segment $[a,b]$ and to adopt the quantization in the
interval strategy.\cite{valent,klauder,gieres} One still employs
the operator $-i\hbar d/dx$, but its domain is required to belong
to $L^2([a,b])$; $a$ and $b$ are the boundary points of the
well. Then, the resulting ``momentum observable" has a discrete
spectrum and the momentum space formulation is given in terms of a
Fourier series.\cite{valent}

Although we arrive at the one-parameter family of momentum-like
operators, the problem is that none of them is
compatible with the infinite well (Dirichlet) boundary conditions.
There is no self-adjoint operator acting as $-i\hbar d/dx$
in the subspace of wave functions in $L^2([a,b])$ which vanish at
the endpoints of the interval.

On the other hand, we should notice that the canonical operators
$X$ and $P$ are defined in $L^2(R)$ without any reference to the
dynamics. Therefore, as a matter of principle, they retain their
physical meaning for any conceivable motion of a particle,
including the permanent trapping conditions. Implicitly, this
viewpoint is represented in Refs.~\onlinecite{cohen},
\onlinecite{robinett}, \onlinecite{doncheski}, and
\onlinecite{majernik}, where $-i\hbar d/dx$ is interpreted in
$L^2(R)$ and is not confined to the interval $[a,b]\subset R$.
Therefore the exterior of the infinite well does matter. The
traditional momentum-space formulation for wave packets,
introduced by the Fourier transform
\begin{equation}
\phi (p,t) = {\frac{1}{\sqrt{2\pi \hbar}}}\int_{-\infty}^{+\infty}
\exp \big(- {\frac{ipx}{\hbar}}\big) \psi(x,t) dx, \label{one}
\end{equation}
 has been exploited in the analysis of the infinite well
and half-line versions of the wave packet
dynamics.\cite{robinett,cohen,doncheski} The notion of a standard
momentum observable with a continuous spectrum also is present in
the derivation of so-called entropic uncertainty relations for the
infinite well.\cite{majernik}

The problem is that the differential expression $-{\hbar^2\over
{2m}}{d^2\over {dx^2}}$, whose domain contains functions $f\in
L^2(R)$ such that $f(x)=0$ if $x\leq a$ and $x\geq b$, is
not a self-adjoint operator in $L^2(R)$. Hence, the infinite
well energy observable definition is defective, if naively
extended to $L^2(R)$ to conform with the presumed domain
properties of $X$ and $P$.

The above mathematical inconsistencies are normally ignored in
the physics oriented literature and the primitive
 (infinite well) example of the quantum mechanical energy spectrum
is not at all analyzed in terms of the full-fledged canonical
quantization formalism. Interestingly, there is also no
agreement among mathematically oriented physicists whether one
can introduce a physically justified candidate for the momentum
operator in the infinite well or the half-line settings. The
folk lore statement reads: there is \it no \rm momentum
observable.\cite{valent}

For the above reasons we reconsider the problem of the quantum dynamics of
a particle that is restricted to a segment of a line by means of impenetrable
barriers. Quantum dynamics with barriers involves a number of
mathematical subtleties: it is necessary to keep in mind the
distinction between symmetric (Hermitian) and self-adjoint
operators. A discussion of self-adjoint extensions of symmetric
operators, with a focus on the teaching of quantum mechanics, can
be found in Ref.~\onlinecite{valent}. Our goal is to resolve the
apparent momentum observable paradoxes\cite{valent,klauder} that prohibit a
consistent use of canonical quantization procedures in the analysis of
quantum systems with trapping boundary conditions.

We resolve the paradox by acknowledging the existence of the rest of
the real line, in conformity with the Fourier transform
definition of Eq.~(\ref{one}), even if we know that the trapped
particle will never occupy that space. The major localization
mechanism is rooted in the dynamics of the particle which is
generated by a properly defined Hamiltonian.

We give physical motivations for the validity of the standard
momentum observable notion for the trapped particle by
investigating the infinite well as the limit of a series of finite
wells. The idealization of an infinite well is given physical
meaning by assuming that it approximately describes more realistic
finite well models. To this end we need to maintain consistent
interpretations of the concepts of position, momentum, and
energy operators in the course of the limiting procedure. This consistency
can be achieved if we consider the infinite well eigenfunctions as the
functions in $L^2(R)$, that is, defined on the whole of $R$, but supported
only by $[a,b]\in R$. We discuss the related energy observable issue in
Secs.~III, IV, and VB. We employ the usual notions of position and
momentum on $R$ and no recourse to momentum-like operators with a
discrete spectrum is necessary.\cite{klauder,valent}

The structure of the paper is as follows. In Sec.~II we outline
the paradoxes that have been found to hamper a consistent
discussion of quantum systems with rigid walls. In Sec.~III we
describe the outcome of a rigorous quantization of particle
motion in a finite interval on the line $R$. In Sec.~IV we
analyze an infinite well as a limit of a finite one and discuss
the groundwork for Sec.~V where we propose to relax the
assumptions of Sec.~III (quantum mechanics in a trap only) by
considering the trap exterior as a necessary element of the
theory. In view of the existence of the standard notions of the
position and momentum observables in $L^2(R)$, the canonical
quantization procedure in the presence of impenetrable barriers is
justified and removes the conceptual obstacles discussed in
Sec.~II.

\section{Quantum systems with barriers -- mathematics versus physics}
Although it is generally accepted that physics is written in the language
of mathematics, there are disagreements on how much mathematical
background is needed to give a proper description of physical phenomena.

The foundations of quantum mechanics employ both the precision of modern
mathematical language and intuition based on the analysis of
physical phenomena. The major developments in quantum theory and its
ability to successfully describe the microworld are due more to physical
intuition than to the precision of mathematics. This success is one
reason why many physicists neglect sophisticated mathematical arguments.

Although we can regard the correspondence between observables and
self-adjoint operators in Hilbert space as generally accepted, the
precise formulation of the operator domains often is considered an
unnecessary nuisance or mathematical pedantry. However, we argue
that the domain subtleties in the operator analysis carry crucial
physical information and must not be disregarded.

The infinite well is a special case of the class of
quantum billiards, which are models of a quantum particle that
is permanently trapped in a bounded region of arbitrary shape. Their energy
spectra can be established only for relatively planar
($R^2$) confinement regions and suffer from the same momentum
observable ``paradoxes" as the infinite well model. Investigations of
the eigenvalue problem for the Laplacian on a connected and compact domain of
arbitrary shape in $R^2$ with Dirichlet boundary conditions have a
long history. In its full generality it is one of the most
difficult problems in mathematics,\cite{hejhal} but suitably
simplified it is a playground for the study of mesoscopic
systems, quantum dots, and other nanostructures.

For a wide class of Hamiltonians, such as those with bounded
potentials, one observes dispersion of wave packets. Thus, even if
the particle is initially confined within a certain interval on
$R$, there is a nonvanishing probability current through the
interval boundaries.

We are interested in the situation when the quantum dynamics is so
restrictive that a particle once localized cannot be found on
certain parts of the real line at any time. This situation amounts
to saying that there is no tunneling,\cite{karw,olk} or any other
form of quantum mechanical transport between those parts and their
complement on $R$. Simple examples of such circumstances are
provided by introducing impenetrable walls. These walls can be
interpreted as ideal trapping enclosures on $R$. Typical barriers
are externally imposed through suitable, often discontinuous and
more singular, potentials. Less spectacular but important examples
of impenetrability are related to the existence of nodes, nodal
curves or surfaces of the generalized ground state function (see
Refs.~\onlinecite{karw} and \onlinecite{olk}).

The notion of impenetrability does not directly follow from the
canonical quantization procedure. A typical quantization recipe first
presumes that there should be primitive kinematic observables related to
the position and momentum, for example, the self-adjoint position and
momentum operators. It is the (secondary) dynamical observable, the
Hamiltonian of the system, that determines the evolution for the system.
Then
$\psi(x,t)$ ultimately appears as a solution of the partial differential
equation with suitable initial/boundary conditions. Hence,
localization essentially arises due to the dynamics with confining
boundary conditions.

Observables are represented by self-adjoint operators which may be
bounded or unbounded. Obviously, the generator of unitary
dynamics, the Hamiltonian, has to be among them. The
self-adjointness property is required because of the spectral
theorem which, as a general solution of the eigenvalue problem for
a given operator, determines a unique link between a operator and
its family of spectral projections. The projection operators in
turn let us state unambiguous elementary (yes-no) questions about
the properties of a physical system. For example, by using
projection operators we may ask for the probability of locating a
particle in a given interval or to find its momentum within a
certain range.

However, in connection with the notion of an unbounded observable,
there are associated very rigid domain restrictions. We shall
address this point in some detail in Sec.~III. An immediate problem can be
seen if we consider a particle on $R$ and assume that it permanently resides
between two impenetrable barriers (rigid walls), placed at points $a$ and
$b$ in $R$. Clearly, the condition $\psi(x,t) =0$ for all $x \leq a$ and $x
\geq b$ is enforced on the wave function of a particle.

One may think that a Hamiltonian can be simply defined as as the
differential operator $-{\hbar^2\over {2m}}{d^2\over {dx^2}}$,
both inside and outside the impenetrable walls. The point is that
such an apparently natural, globally defined Hamiltonian is not a
self-adjoint operator. It is not even a symmetric
operator.\cite{schechter} Hence, a consistent definition of the
quantum dynamics in the presence of a barrier needs a careful
examination of self-adjoint operator candidates for the
Hamiltonian of the quantum system.

Another obvious conflict with intuition appears when one tries to
interpret the differential expression $-i\hbar d/dx$ as a momentum
operator in the barrier context. The continuous spectrum of the
momentum operator for a free quantum particle on a line is well
known. The notion of momentum is not so obvious for the infinite
well model in view of the textbook wisdom: ``\ldots momentum
operator eigenfunctions do not exist in a box with rigid walls,
because then they would vanish everywhere.''\cite{schiff} In
contrast, another well known textbook\cite{cohen} does not
prohibit such notions as the momentum measurement and the
distribution of continuous momentum values in stationary states,
these being interpreted as $L^2(R)$ wave packets. A quantum
particle in an infinite well gives rise to a pictorial
illustration of the wave packet dynamics.\cite{cohen,doncheski}

An attentive reader must be confused, because both discussions
seem to be justified,\cite{schiff,cohen} although the
discrepancies between the two points of view were not explained or
resolved in a single text. In Refs.~\onlinecite{cohen},
\onlinecite{robinett}, and \onlinecite{doncheski}, an explicit
answer was formulated for the probability of a measurement of the
momentum $P$ of the particle yielding a result between $p$ and
$p+dp$ for a particle confined in an infinite well. All
calculations explicitly involve the $L^2(R)$ Fourier integral
Eq.~(\ref{one}) for spatially confined wave packets, thus
suggesting that the infinite well problem may not be in conflict
with the standard notion of the momentum operator (understood as
the generator of spatial translations in $L^2(R)$). Such an
operator has a continuous spectrum.

The same infinite well problem has been summarized in
Ref.~\onlinecite{klauder} as follows: the spectrum of the operator
$P$ is discrete, hence the Hilbert space in the momentum
representation becomes the Hilbert space $l^2$ of square summable
sequences, see for example, Sec.~III. Then, Eq.~(\ref{one}) is
interpreted as a mathematically equivalent version of the infinite
well wave function $\psi(x,t)$, but \it not \rm as its momentum
representation.

In Refs.~\onlinecite{cohen}, \onlinecite{robinett}, and
\onlinecite{doncheski}, the differential expression $-i\hbar d/dx$
is interpreted in $L^2(R)$, hence the exterior of the infinite
well does matter. In Ref.~\onlinecite{klauder}, the same
differential expression is localized to the interior of the well
by demanding that its domain belongs to $L^2([a,b])$,with $a$ and
$b$ the well boundaries, so the rest of the line is irrelevant.

Analogous conflicting interpretations can be seen in the
discussion of a single impenetrable barrier that divides $R$ into
two non-communicating segments, see for example,
Refs.~\onlinecite{valent} and \onlinecite{doncheski}. A quantum
particle, once initially localized on the half-line, either
positive or negative, would reside on the half-line indefinitely,
with no chance to change the localization area. Again, the usual
momentum representation\cite{cohen,doncheski,robinett} makes sense
in the analysis of the dynamical behavior of wave packets.
However, it is well known\cite{simon} that a symmetric operator
$-i\hbar {\partial \over {\partial x}}$, as defined on
$C_0^{\infty}(R^{\pm})$ (the space of the infinitely
differentiable functions of compact support in the positive $R^+$
or negative $R^-$ half-lines of $R$), has no self-adjoint
extensions in $L^2(R^+)$ or $L^2(R^-)$. In other words there is no
self-adjoint momentum operator of the form $-i\hbar {\partial
\over {\partial x}}$ for a particle on a half-line. Accordingly,
the authors of Ref.~\onlinecite{valent} conclude that ``\ldots the
momentum is not a measurable quantity in that situation.''

To summarize, the standard Fourier integral analysis on the real
line, Eq.~(\ref{one}), has been applied to wave packets of a
particle confined to a segment of $R$ or to the half-line and
interpreted as a consistent spectral analysis of the momentum
operator.\cite{robinett,cohen,doncheski} According to
Refs.~\onlinecite{valent} and \onlinecite{klauder}, the previous
analysis can be seen only as an admissible computational device
having nothing to do with the momentum operator and the true
physically relevant state of affairs for a particle
confined to the segment is said to refer to the spectral
analysis of the momentum operator in terms of Fourier series. For
a particle confined to the half-line, the notion of momentum is
said not to be defined.

\section{Quantization in the finite interval}
We now discuss the mathematical issues of the quantization on
the interval (a particle confined to a segment of $R$). We begin
with some observations concerning a free particle on the real line
$R$.

In one-dimensional models on the real line, the momentum operator
$P$ and the free Hamiltonian $H$ are self-adjoint operators
defined by $-i\hbar d/dx$ and $(- \hbar^2/2m) d^2/dx^2$
respectively. However, these standard differential expressions,
when defined on the space $C_0^{\infty}(R)$ of infinitely
differentiable functions of compact support, are not self-adjoint
but only symmetric operators. In the following, all coefficients
such as $\hbar$ and $\hbar^2/2m$ will be set equal to unity for
convenience.

Because $C_0^{\infty}(R)$ is invariant under differentiation, the
symmetric operator $-{{d^2}\over {dx^2}}$ can be interpreted as
the square of another symmetric operator $-i{{d}\over {dx}}$, in
the sense that it means two consecutive actions. To obtain the
self-adjoint operators from the symmetric ones, we must expand
their domains. There are a priori two possibilities:

\begin{enumerate}[(i)]

\item We can extend the symmetric operator $-i{{d}\over {dx}}$ by
taking its closure to a self-adjoint operator $P$, which is then
called a momentum operator, and define the free particle
Hamiltonian operator $H_f= P^2$.

\item We can extend the symmetric operator $-{{d^2}\over {dx^2}}$ by
taking its closure to a self-adjoint operator $\tilde{H}_f$ which
may be called the Hamiltonian operator.

\end{enumerate}
These two procedures give the same result: $H_f = P^2 =
\tilde{H}_f$ if considered in $L^2(R)$.

The situation is different when we pass to $L^2([a,b])$, because
now the mathematical subtleties unavoidably enter. It turns out
that there is not one, but a family of infinitely many
self-adjoint operators in $L^2([a,b])$ whose action on functions
from the domain is defined by the same expression
$-i{\frac{d}{dx}}$. In the following, we shall simplify the
notation by choosing $a=0$, $b=\pi$ and hence the Hilbert space
$L^2([0,\pi])$.

The differential expressions $-i{{d}\over {dx}}$ and\
$-{{d^2}\over {dx^2}}$ when acting in $C_0^{\infty}(0,\pi)$
(infinitely differentiable functions with support included in the
open interval $(0,\pi) \subset R$) define symmetric operators in
$L^2([0,\pi])$. Obviously, $C_0^{\infty}(0,\pi)$ is invariant
under differentiation and thus $-{{d^2}\over {dx^2}}$ is the
square of $-i{{d}\over {dx}}$ in the sense of two consecutive
actions. However, now the procedures (i) and (ii) require some
care. In what follows we shall refer to the Krein-von Neumann
theory of self-adjoint extensions, see for example
Refs.~\onlinecite{valent}, \onlinecite{glazman}, and the Appendix.

Let us begin with procedure (i). We denote $A= -i{{d}\over {dx}}$ on
$C_0^{\infty}(0,\pi)$. Then its closure $\overline{A}= -i{{d}\over {dx}}$ is
defined as the differential expression
$-i{{d}\over {dx}}$ acting on an expanded domain
$D(\overline{A}) = \{ f \in AC[0,\pi]; f(0)=0=f(\pi)\}$. The notation $AC$
refers to the absolute continuity of $f$ which gives meaning to the first
derivative $f'$. The boundary conditions emerge in the process of taking the
closure.

The operator $\overline{A}$ is a closed symmetric operator, but is
not self-adjoint. To find the self-adjoint extension of
$\overline{A}$, we need to establish its deficiency
indices.\cite{valent,glazman,simon} In the Appendix we show them
to be $(1,1)$, which implies that $\overline{A}$ has a one
parameter family of self-adjoint extensions in $L^2([0,\pi])$. We
denote the extensions by $P_{\alpha}$:
\begin{equation}
P_{\alpha} = -i{{d}\over {dx}} \quad
D(P_{\alpha}) = \{ f \in AC[0,\pi]; f(0)= \exp(i\alpha)
f(\pi) \} \qquad (0 \leq \alpha < 2\pi) . \label{two}
\end{equation}

Note that there are no other self-adjoint extensions of
$\overline{A}$, and thus no other self-adjoint operators acting as
$-i{{d}\over {dx}}$. For each $\alpha$, there is in
$L^2([0,\pi])$ an orthonormal basis that is composed of the
eigenvectors of $P_{\alpha}$,
\begin{equation}
e^{\alpha}_n(x) = {1\over \sqrt{\pi}} \exp i(2n+{\alpha \over
\pi})x, \label{three}
\end{equation}
where $n$ takes integer values, and the eigenvalues of $P_{\alpha}$
are
\begin{equation}
p_n^{\alpha} = 2n + {{\alpha}\over {\pi}}. \label{four}
\end{equation}

Let us introduce another definition for $D(P_{\alpha})$. If $f\in
L^2([0,\pi])$ is expressed in terms of $e^{\alpha}_n$ so that
$f(x) = \sum_n f^{\alpha}_n e^{\alpha}_n(x)$, then $f\in
D(P_{\alpha})$ if and only if $ \sum_n n^2 |f^{\alpha}_n|^2 <
\infty$. This supplementary characterization of the domain will
prove useful to define functions of the operators $P_{\alpha }$,
c.f.\ the spectral theorem description in the Appendix.

The operator $H_{\alpha}$ defined by
\begin{equation} H_{\alpha} =
(P_{\alpha})^2, \label{seven}
\end{equation}
 has the same family of eigenvectors as $P_{\alpha}$, but its
eigenvalues are
\begin{equation}
E^{\alpha}_n = (p^{\alpha}_n)^2 = (2n + {\alpha \over \pi})^2
\label{8}
\end{equation}
for all integers $n$. As a consequence,
\begin{equation}
D(H_{\alpha}) = \{ f = \sum_n f^{\alpha}_n e^{\alpha}_n ; \sum_n
n^4 |f^{\alpha}_n|^2 < \infty \}. \label{domain}
\end{equation}
Thus $D(H_{\alpha}) \subset D(P_{\alpha})$ and $D(P_{\alpha}) =
P_{\alpha} D(H_{\alpha})$.
It also follows that
\begin{equation}
H_{\alpha} = -{{d^2}\over {dx^2}} \quad
D(H_{\alpha}) = \{ f\in AC^2[0,\pi]; f(0)= \exp (i\alpha)
f(\pi), f'(0) = \exp(i\alpha) f'(\pi)\}, \label{ten}
\end{equation}
where the $AC^2$ notation gives meaning to the second
derivative of $f$.
Therefore the operator $H_{\alpha}$ in Eq.~(\ref{ten}) can be safely
interpreted as two consecutive actions of $P_{\alpha}$, Eqs.~(\ref{two})
and (\ref{seven}), where both operators are self-adjoint.

Now let us consider (ii). The closure of $-{{d^2}\over {dx^2}}$ as
defined on $C_0^{\infty}(0,\pi)$ is $\overline{H} = -{{d^2}\over
{dx^2}}$ with the domain $D(\overline{H}) = \{f\in AC^2[0,\pi];
f(0) = f(\pi)= f'(0)= f'(\pi) = 0\}$. The closed symmetric
operator $\overline{H}$ has the deficiency indices $(2,2)$.
Therefore the family of all self-adjoint extensions of
$\overline{H}$ is in one-to-one correspondence with $U(2)$, the
family of all $2\times 2$ unitary matrices, see for example,
Refs.~\onlinecite{valent} and \onlinecite{glazman}.

We can devise a family of $U_{\alpha}\in U(2)$, $0\leq \alpha
<2\pi$, whose choice is equivalent to the boundary conditions
$f(0) = \exp(i\alpha) f(\pi)$, $f'(0) = \exp(i\alpha) f'(\pi)$,
and thus defines $H_{U_{\alpha}} = H_{\alpha}$, Eq.~(\ref{seven}),
with the domain $D(H_{\alpha})$, Eq.~(\ref{ten}). Consequently,
the two procedures (i) and (ii), are equivalent for all operator
pairs $H_{\alpha}$, $P_{\alpha}$ with $0\leq \alpha < 2\pi$.

The family $U_{\alpha}$ is a proper subset of $U(2)$ and thus
there are $H_U$ for which (i) does not work. For example, for a
suitable choice of a unitary matrix $U$,\cite{valent} the
corresponding self-adjoint operator $H_U \doteq H_w$ is the
infinite well Hamiltonian:
\begin{equation}
(H_wf)(x) = - {{d^2}\over {dx^2}} f(x) \quad D(H_w) = \{ f\in
AC^2[0,\pi]; f(0)=f(\pi) = 0 \} . \label{11}
\end{equation}

In the infinite well context provided by Eq.~(\ref{11}), we are
not allowed to interpret $H_w$ as the square of any self-adjoint
$-i{\frac{d}{dx}}=P_{\alpha}$. The reason is that no $P_{\alpha}$
respects the Dirichlet boundary condition, which makes it
impossible to identify the Hamiltonian $H_w$ in $L^2([0,\pi])$ as
$P^2_{\alpha}$. Consequently, the quantization in a finite
interval gives rise to:

\begin{enumerate}[(i)]

\item The one-parameter family of Hamiltonians $H_{\alpha}$ of Eq.~(\ref{ten})
with the momentum operators $P_{\alpha}$ of Eq.~(\ref{seven}),
whose eigenvalues form discrete spectra,

\item The Hamiltonian $H_w$ of Eq.~(\ref{11}), suitable for the infinite
well problem, but then with no notion of a momentum observable.

\end{enumerate}

To
complete the quantization scheme on the interval, we need to introduce
the position operator $Q$ defined as
$(Qf)(x)=xf(x)$. In the present case it is a bounded operator, contrary
to what is normally expected from a member of a canonically conjugate
position-momentum pair.

The canonical commutation relations
$QP_{\alpha} - P_{\alpha}Q = iI$ formally hold on all $f\in AC(a,b);\,
f(a)=f(b)=0$, but cannot be given in Weyl form (that is, in terms of
suitable unitary operators) which is indispensable for the mathematical
consistency of the canonical formalism. Note that by following the
procedure (i), which yields Eq.~(\ref{seven}), we have lost a direct
link to the infinite well problem.

For the special case of $\alpha =0$, we end up with a degenerate
spectrum $E_n = (2n)^2$. This spectrum corresponds to the familiar
plane rotator. For $\alpha \neq 0$, we can relate the spectral
problem Eq.~(\ref{8}) to the rotation of a charged particle around
an infinitely thin solenoid;\cite{carlen} the parameter $\alpha$
is related to the magnetic flux. Hence, $H_{\alpha}$, $P_{\alpha}$
refer exclusively to rotational (angular dynamics) features of
motion. Neither ${\frac{d^2}{dx^2}}$ with the Dirichlet boundary
condition, nor any other $H_U$ (provided $U\neq U_{\alpha}$) fit
the above canonical quantization picture; we recall that no
self-adjoint momentum operator of the form $-i\frac{d}{dx}$
is compatible with the Dirichlet boundary conditions.

In connection with Eq.~(\ref{8}), the textbook solution of the infinite well
yields the familiar spectral formula
$E_n = n^2$, where $n\geq 1$ is a natural number. This result is
incompatible with $E^{\alpha}_n = (2n + {\alpha \over \pi})^2$,
Eq.~(\ref{8}) where $n$ is an integer. Moreover, the related eigenfunctions
$e^{\alpha}_n(x)$, do not respect the Dirichlet boundary conditions in
contrast to the ``true'' infinite well Hamiltonian eigenfunctions $\psi_n(x)
= \sqrt{2\over \pi} \sin nx$. A possible physical interpretation of $H_U$
that falls neither in the class (\ref{ten}) nor (\ref{11}) is discussed in
Ref.~\onlinecite{valent}.

We stress that the interpretation of $P_{\alpha}$ in
Eq.~(\ref{two}) as a momentum operator for a trapped particle (as
advocated in Refs.~\onlinecite{valent,gori,klauder}) stems from
the fact that its differential expression reads
$-i{\frac{d}{dx}}$, just as it does for a particle on the real
line. Some obvious consequences of this implicit $L^2(R)$ input in
the isolated trap, $L^2([a,b])$, include: (1) the non-uniqueness
of the momentum operator, (2) the non-existence of the momentum
operator on the half-line, and (3) a conceptual discontinuity in
the interpretation of the momentum observable between $L^2(R)$ and
$L^2([a,b])$ $L^2([a,\infty])$.\cite{valent} The latter conceptual
discontinuity relates to the limiting procedures when passing from
regular (such as the finite well with its unique momentum
observable) to singular problems (such as the infinite well, or
half-line cases, with non-unique or no momentum observable).

\section{The infinite well as the limit of the finite ones}

It is common for physicists to replace a complicated physical
system by a simpler solvable model and then obtain approximate
answers to the originally posed questions. Often the solvable models are
more singular than the realistic ones. In quantum mechanics textbooks,
the piecewise constant potentials that form sharp barriers, steps or
wells are implicitly interpreted as idealized versions of continuous
potentials of similar shapes. A more singular example is the Dirac delta
potential which often is used instead of a very narrow and very deep
potential well.\cite{karw,holden}

Infinite well (or infinite barrier) models make sense if they
are capable of giving approximate answers to questions concerning finite
wells. It is important that the validity of the approximation be
controlled, which requires the notion of continuity when passing from
the finite well to the infinite one. In this section, we are
motivated by the considerations of Ref.~\onlinecite{valent} where the
previously mentioned conceptual discontinuity between the finite well
and infinite well models is clearly emphasized.

It is natural to consider the half-line case as the limit of the
step potential. Again we encounter problems with the idea of the momentum
observable: for any finite height of the step potential, there
exists a momentum observable (a unique self-adjoint operator acting as
the differential expression $-i\hbar {\frac{d}{dx}}$), while for an
infinite height there is no self-adjoint extension corresponding to
$-i\hbar {\frac{d}{dx}}$. The conclusion of Ref.~\onlinecite{valent}
(see Sec.~7.4), that ``an infinite potential cannot be simply described
by the limit of a finite one'' contributes to the paradoxes
and inconsistencies we discussed in Sec.~II.

 If one tries to model a
particle that is localized on a segment of a line, the confinement
is enforced by considering Hamiltonians with vanishing boundary
conditions at the ends of the interval. This boundary condition
can be imposed either by the singularity of the potential (such as
the P\"{o}schl-Teller potential in Ref.~\onlinecite{klauder}) or
``by hand'' as for the infinite well.\cite{klauder,valent} The
latter case is justified by introducing the vague concept of a
finite potential within the spatial segment and plus infinity
otherwise.

The reasoning goes as follows. A particle that is trapped inside
the infinite well $0\leq x\leq \pi$ must have its wave function
equal to zero outside the well. To ensure this condition, we
consider the potential $V(x) = \infty$ on the complement of the
open interval $(0,\pi)$ in $R$, while $V(x) = 0$ between the
impenetrable barriers.

Note that the corresponding
stationary Schr\"{o}dinger equation,
\begin{equation}
[- \nabla^2 + V(x)] \psi(x) = E \psi(x), \label{10}
\end{equation}
with $x\in R$ has no meaning beyond the chosen interval.

By formally setting $\infty \times 0 = 0$ in the ``improper''
area, one argues that in view of Eq.~({\ref{10}), the wave function
$\psi (x)$ must vanish for $x\leq 0$ and $x\geq \pi$.
 Then, one concludes that instead of demanding
unusual properties of $V(x)$, it is more natural to impose
restrictions on the wave functions demanding that $\psi\in
L^2([0,\pi])$; $\psi(0) =\psi(\pi)=0$ (the dynamics is spatially
restricted to $[0,\pi]$). In other words, the rest of the line can
be neglected.

Now, let us consider a (dis)continuity in passing to the
infinite well from a finite well. We have mentioned that the
infinite well problem acquires a physical meaning as an
approximation (by suitable limiting procedures) of a finite well
model.
Let us consider\cite{valent} $V(x) = 0$ for $x\in (0,\pi)$
and $V(x) =V_0
>0$ for $x \notin (0,\pi)$. As $V_0 \to \infty$, the number of
eigenvectors for the finite well problem $- \nabla^2 + V$ also
goes to infinity. Let us label by $n \in N$ the eigenvalues
$E^V_n$ in increasing order and the corresponding eigenfunctions
by $\phi_n^V$:
\begin{equation}
(- \nabla^2 + V)\phi^V_n = E^V_n \phi^V_n.
\end{equation}

For fixed $n$ we obtain for large values of $V_0$ (compare for
example, Ref.~\onlinecite{valent}):
\begin{equation}
E^V_n \simeq E^{\infty}_n (1 - {\frac{4}{\pi
\sqrt{V_0}}}), \label{14}
\end{equation}
where $E^{\infty}_n = n^2$ is the infinite well
energy eigenvalue with $n=1,2,\ldots$ We also have
\begin{subequations}
\label{15}
\begin{eqnarray}
\phi^V_n(x\leq 0) &\simeq& \sqrt{{\frac{2}{\pi}}}
\left({\frac{n}{\sqrt{V_0}}} \right)
\exp\{ - |x| \sqrt{V_0} \} \\
\phi^V_n(0\leq x\leq \pi) &\simeq& \sqrt{{\frac{2}{\pi}}}
\big[\sin nx + ({\frac{1}{\pi \sqrt{V_0}}}) \left[(n\pi) \cos nx - \sin
nx\right] \big] \\
\phi^V_n(x\geq \pi) &\simeq& \pm \sqrt{{\frac{2}{\pi}}}
\big({\frac{n}{\sqrt{V_0}}} \big)
\exp[- (x- \pi) \sqrt{V_0}].
\end{eqnarray}
\end{subequations}
Accordingly, when $V_0 \rightarrow \infty$, then $E^V_n \rightarrow
E^{\infty}_n$, and
\begin{equation}
\phi^V_n(x) \rightarrow \phi^{\infty}_n(x) = \sqrt{{\frac{2}{\pi}}}
\sin nx \label{16}
\end{equation} for $0\leq x\leq \pi $ and zero otherwise. The infinite
well Hamiltonian eigenvalues and eigenfunctions are thus smoothly
reproduced and we keep under control the accuracy of the approximation
of the finite well by its infinite well idealization.

We need to achieve more than the convergence properties
Eqs.~(\ref{14}}) and (\ref{16}). Namely, we are interested in
verifying whether the finite well notions of position, momentum,
and energy observables go through the limiting procedure. (We
recall the no-go claim of Ref.~\onlinecite{valent}.)

Note that the limit $\phi^V_n
\rightarrow \phi^{\infty}_n$ as $V_0 \rightarrow \infty$ holds in
the norm of $L^2(R)$. It follows that for any interval
$(x_1,x_2)$, we have, using an obvious notation, the following
behavior of the localization probabilities: $P^V_{x \in (x_1,x_2)}
\doteq \int_{x_1}^{x_2} |\phi^V_n(x)|^2 dx \rightarrow
\int_{x_1}^{x_2} |\phi^{\infty}_n(x)|^2 dx = P^{\infty}_{x\in
(x_1,x_2)}$ as $V_0 \rightarrow \infty$. So, we have secured the
standard meaning of the position measurement for both the
finite and infinite well problems.

These limiting behaviors are paralleled by the convergence of the
suitable Fourier transforms. Indeed, it is well known that the
Fourier transform, as defined in $C_0^{\infty}(R)$, can be
extended to a unitary operator in $L^2(R)$. Therefore, the Fourier
transform of $\phi^V_n$ also converges in the $L^2(R)$ norm to the
Fourier transform ${\cal{F}}\phi^{\infty}_n$ of $\phi^{\infty}_n$.
Hence, for any $(p_1,p_2)$, we have that $P^V_{p \in (p_1,p_2)}
\doteq \int_{p_1}^{p_2} |{\cal{F}}\phi^V_n(p)|^2 dp \rightarrow
\int_{p_1}^{p_2} |{\cal{F}}\phi^{\infty}_n(p)|^2 dp =
P^{\infty}_{p\in (p_1,p_2)}$ as $V_0 \rightarrow \infty$. Thus, we
conclude that if the infinite well problem eigenfunctions are
considered as the functions defined on $R$ but supported by
$[0,\pi]$, then we can employ the usual notions of position and
momentum on $R$ and these notions are common for the finite and
the infinite well. The conceptual continuity in the notions of
position, momentum, and energy measurements survives the limiting
procedure $V_0 \rightarrow \infty$.

We emphasize that for $L^2(0,\pi)$, we have two nonequivalent ways
of making the Fourier analysis. If $L^2(0,\pi)$ is considered as a
subspace of $L^2(R)$, then ${\cal{F}}L^2(0,\pi) \subset L^2(R)$.
More precisely, if $0\neq f \in L^2(0,\pi)$, then ${\cal{F}}f \in
L^2(R)$, but ${\cal{F}}f$ does not belong to $L^2(0,\pi)$. Because
the support of $f$ is compact, the function ${\cal{F}}f$ can be
analytically continued to the entire complex plane. Thus, if
${\cal{F}}f$ vanishes on $R\backslash [0,\pi]$, it also vanishes
identically on $R$.

If $R\backslash [0,\pi]$ is neglected and $L^2(0,\pi)$ is
considered independently, then we can employ the Fourier series.
In the language of Ref.~\onlinecite{klauder}, the Fourier series
stands for the momentum representation formulation if the momentum
operator is chosen to be $P_0$, as given by Eq.~(\ref{two}). The
Hilbert space of this momentum representation is then $l^2(Z)$,
the space of square summable sequences $f_n$, where $n$ runs over
the set of integers $Z$. Let us note that the self-adjoint
operators, $P$ in $L^2(R)$ and $P_0$ in $L^2(0,\pi)$, both
exemplify the spectral theorem and the notion of momentum
representation, but are fundamentally different operators.

In the course of all limiting operations, the notion of $L^2(R)$
and thus of the entire real line input (notably of the usual
momentum observable) is implicit. This observation lends support
to the standard momentum representation concept, employed in
Refs.~\onlinecite{robinett}, \onlinecite{cohen}, and
\onlinecite{doncheski}, which can thus be adopted to the infinite
well and the half-line wave packet dynamics. Consequently, if we
had followed the strategy of Refs.~\onlinecite{klauder},
\onlinecite{valent}, and \onlinecite{gieres} and ignored the rest
of the real line, the restriction of the model to $L^2([0,\pi])$
would have ruled out ${\cal{F}}\phi^{\infty}_n$. As a result, the
usual concept of the momentum operator as the generator of the
translation group would no longer be appropriate and the
interpretation in Ref.~\onlinecite{valent} would make a sharp
distinction between the finite well and infinite well cases. Such
a distinction is untenable on physical grounds.

\section{Quantum dynamics with barriers}

\subsection{Trapping as a dynamical effect}

Now we shall analyze the main outcome of our
previous discussion: we can make sense of various operators for
trapped particles by not ignoring the rest of the real line
(the exterior of the trap).

In the canonical quantization scheme, quantum mechanics on the
entire real line refers to the correspondence principle, which
introduces the position $Q$ and momentum $P$ observables as
unbounded operators in $L^2(R)$. The intuitive definition of
multiplication and differentiation operators on smooth functions
with a reasonable fall off at infinity is sufficient to determine
uniquely the conjugate self-adjoint operators that obey the
canonical commutation relations in the Weyl form (that is, by
means of unitary operators). This statement is purely kinematical
and thus independent of any dynamics.

The free particle Hamiltonian,
\begin{equation}
H_f = - {\frac{d^2}{dx^2}} = P^2,\label{18}
\end{equation}
implies that $P$ commutes with $H_f$, and thus is a constant of
motion which supports the view that $P$ is the momentum operator. For the
free particle the identity (\ref{18}) relates the Hamiltonian $H_f$ and
$P^2$. In other cases, there appear potentials or boundary conditions (such
is the case for the half-line and infinite well problems). Whatever the
dynamics and thus the general Hamiltonian $H$ may be, we can safely assume
that $H$ is self-adjoint and bounded from below.

Let us consider the general mathematical mechanism of permanent
confinement. Let $H$ be a Hamiltonian operator and we choose an
open interval $G\subset R$ with $\chi_G$ denoting its
characteristic (indicator) function: $\chi_G(x) =1$ for $x\in G$
and vanishes otherwise. (To conform with the previous notation, we
suggest the identification $G\doteq (a,b)$ and $\overline{G}
\doteq [a,b]$.)

If $f\in D(H)$, then $\chi_Gf$ typically does not belong to
$D(H)$. If, however, for a given $H$ and $G$, the property $f \in
D(H)$ necessarily implies that $\chi_Gf\in D(H)$ then $\chi_G$,
considered as a projection operator in $L^2(R)$, commutes with the
spectral projectors of $H$ and hence with the unitary operator
$\exp(-iHt)$. This property implies an invariance of the subspace
$[f\in L^2(R); {\rm supp}\, f \subset \overline{G}]$ with respect
to time evolution. Thus, if at some instant of time a particle is
localized in $\overline{G}$, that is, its wave function $f$ is
supported by a subset of $\overline{G}$, then $supp \{ g(t)=
\exp(-iHt) f\} \subset \overline{G}$ for all times $t$. Hence the
particle has always been in $\overline{G}$ and will stay there
forever.

Consequently, if the dynamics is defined by the Hamiltonian $H$ in
$L^2(R)$, then the confinement in $\overline{G}$ occurs if and
only if $H$ can be split into a direct sum $H= H_1 \bigoplus H_2$
corresponding to the decomposition $L^2(R) = L^2(R\backslash G)
\bigoplus L^2(\overline{G})$, so that $H_1$ is self-adjoint in
$L^2(R\backslash G)$ and $H_2$ is self-adjoint in
$L^2(\overline{G})$. Then $\exp(-iH_1t)$ and $\exp(-iH_2t)$
describe the time evolution of the system localized in $R
\setminus G$ and $\overline{G}$ respectively. Moreover $\exp(-iHt)
= \exp(-iH_1t)\exp(-iH_2t)$.

Thus the dynamics from the outset takes account of the
impenetrable barrier at the boundary of $G$. This effect is purely
dynamical, and there is no reason to modify the meaning of
kinematical variables such as the position and momentum (see
Sec.~III). Consequently, if a particle described by the wave
function $f(x)$ is localized in $\overline{G}$, then necessarily
$f\in L^2(\overline{G})$. However, now the momentum representation
reads $f(x) \rightarrow ({\cal{F}}f)(p)\doteq \tilde{f}(p) $, by
the Fourier integral, Eq.~(\ref{one}). If $G$ is bounded, then
${\cal{F}}f$ is an entire function. So, if a particle at some
(initial) instant of time is localized in a bounded region in
space, then its momentum is spread over the whole real line.

In the following we illustrate the qualitative physical and
mathematical mechanisms leading to the above reduction of $L^2(R)$
by the dynamics.

\subsection{Infinite well}
First, we define $H= - {{d^2}\over {dx^2}}$ through its specific
domain $D(H) = [f\in AC^2(R); f, f', f'' \in L^2(R), f(0) =0=
f(\pi)]$. We recall that the $AC^2$ notation refers to the
absolute continuity of the first derivative which gives meaning to
the second derivative (in the sense of distributions, as a
measurable function). The operator $\{ H, D(H)\} $ is self-adjoint
and the decomposition $L^2(R) = L^2(R\backslash G) \bigoplus
L^2(\overline{G})$, together with $H= H_1 \bigoplus H_2$, holds
for $\overline{G}=[0,\pi]$. Thus the traditional infinite well
problem is nothing else than the analysis of $H_2$ in the space
$L^2([0,\pi])$, with the Dirichlet boundary condition. Here, $H_2
= H_w$, see for example Eq.~(\ref{11}).

\subsection{Centrifugal repulsion}

Let us consider the operators belonging to the family of singular problems
with the centrifugal potential (possibly modified by harmonic
attraction)\cite{calogero,olk}:
\begin{equation} H = - {d^2 \over {dx^2}} + {1\over {[n(n-1)x^2]}},
\label{19}
\end{equation}
with $n\geq 2$ and $D(H) = [f\in AC^2(R); f, f', f" \in L^2(R),
f(0)=0=f'(0)]$. The operator $H$ in Eq.~(\ref{19}) is
self-adjoint. The projection operator $P_+$ defined by $(P_+
f)(x)= \chi_{R^+}(x) f(x)$ clearly commutes with $H$. The
singularity of the potential is sufficiently severe to enforce the
boundary condition $f(0) =0=f'(0)$ (the generalized ground state
function (cf.~Ref.~\onlinecite{berezanski}) may be chosen for this
scattering problem in the form $\phi (x)= x^n$).

The Hilbert spaces $L^2(R^+)$ and $L^2(R^-)$ are invariant under
the Schr\"{o}dinger evolution $\exp(-iHt)$ generated by $H$ and
the Schr\"{o}dinger probability current vanishes at $x=0$ for all
times. Consequently, there is no dynamically implemented
communication between the two disjoint localization areas
extending to the negative or positive semi-axes of $R$
respectively. The respective localization probabilities of finding
a particle on a positive or negative semi-axis are constants of
the motion. Because of the singularity at $0$, once trapped, a
particle is confined in one particular enclosure only and cannot
be detected in another.

However, we note that $D(H)$ contains functions from $L^2(R)$ that
are restricted to obey $f(0)=0=f'(0)$ and not necessarily to
vanish on either half-line. Such functions may have support on
both the positive and negative semi-axes simultaneously. For
example, a normalized linear combination of two components
corresponding to positive and negative half-lines respectively, is
a legitimate element of $D(H)$. Then, we can merely predict a
probability to detect a particle on either side of the origin.
This probability is a constant of the motion, and there is no
probability current through the origin. In particular, due to the
boundary conditions, if $f\in D(H)$ then $\chi _+f \in D(H)$ and
$\chi _-f \in D(H)$.

The classic Calogero-type problem is defined by
\begin{equation} H = - {d^2\over {dx^2}} + x^2 + {\gamma \over
{x^2}}. \label{twenty}
\end{equation}
The eigenvalues
are $E_n = 4n + 2 + (1+4\gamma)^{1/2}$, where $n\geq 0$ and $\gamma >
-{1\over 4}$, with eigenfunctions of the form:
\begin{eqnarray}
f_n(x) &=& x^{(2\alpha +1)/2} \exp(-{x^2\over 2})\,
L^{\alpha}_n(x^2)\\
L_n^{\alpha} (x^2)& =& \sum_{\nu =0}^n {{(n+\alpha)!}\over {(n-\nu)!
(\alpha + \nu)!}} {{(-x^2)^{\nu}}\over {\nu !}},
\end{eqnarray}
where $\alpha = {1\over 2}(1+4\gamma)^{1/2}$. The $\gamma$
parameter range, $-1/4< \gamma < 3/4$, involves some mathematical
subtleties concerning the singularity at 0 that are not
sufficiently severe to enforce the Dirichlet boundary
condition.\cite{falomir,simon} However, in the range $\gamma \geq
3/4$ the ground state is doubly degenerate in the whole
eigenspace of the self-adjoint operator $H$. The singularity at
$x=0$ decouples $(-\infty,0)$ from $(0,+\infty)$ so that
$L^2(-\infty,0)$ and $L^2(0,+\infty)$ are the invariant subspaces
for the dynamics generated by $H$.

The singularity in both Hamiltonians (\ref{19}) and
(\ref{twenty}) can be removed by a simple replacement $x^2
\rightarrow (x^2 + \epsilon)$ with $\epsilon >0$. The limit
$\epsilon \rightarrow 0$ would restore the singularity. As with
 the infinite well limit for finite wells, the relatively easy
to solve singular models (\ref{19}) and (\ref{twenty}) may be
considered as approximations of more complicated regular (free of
singularities) models.

We emphasize that impenetrable barriers are located at points
where a potential singularity enforces vanishing boundary
conditions. In particular, such conditions are satisfied by
(generalized) ground states and this mathematical feature is
responsible for the appearance of impenetrable barriers. Let $\phi
\in L^2_{\rm loc}(R)$, that is, we consider all functions that are
square integrable on all compact sets in $R$. If there is a closed
set $N$ of Lebesgue measure zero so that (strictly speaking we
admit distributions) ${{d\phi}\over {dx}} \in L^2_{loc}(R
\setminus N)$, then there is a uniquely determined Hamiltonian $H$
such that $\phi$ is its (generalized) ground state. If
$(x-x_0)^{-1/2} \phi $ is bounded in the neighborhood of $x_0$,
then there is an impenetrable barrier at $x_0$. For a precise
description of this mechanism in $R^n$, see for example,
Ref.~\onlinecite{karw}.

\subsection{Multi-trapping enclosure}

In contrast to the centrifugal repulsion where the singularity of
the potential alone was capable of making the ground state
degenerate due to the impenetrable barrier at the origin, we also
can impose the existence of barriers as an external boundary
condition. We introduce the differential expression $H_0= - {d^2
\over {dx^2}}$ and observe that for any real $q$, the function
$\psi(x)= \sin(qx)$ satisfies the equation $H_0\psi= q^2\psi$. The
operator $H_q= H_0 - q^2 $ is self-adjoint when operating on
$D(H_q)= [f\in AC^2(R); f, f', f'' \in L^2(R), f({{n\pi}\over
q})=0, n=0, \pm 1, \pm 2,\ldots]$ and $\sin(qx)$ is its
generalized ground state. In this case a particle localized at
time 0 in a segment $[(n-1){\pi \over q}, n{\pi \over q}]$ will be
confined there forever. This model can be considered as that of
multi-trapping enclosures, with impenetrable barriers at points
$n{\pi \over q}$. Note that in every segment $[{(n-1){\pi  \over q}},
n{{ \pi} \over q}]$, the corresponding dynamics is identical with the
one associated previously with the infinite well.

\section{Conclusion}
We have considered several singular models (such as the infinite
well) that serve as approximations of regular ones (such as the
finite well) in the sense of suitable limits. If the properties
of the limiting model are to give a reliable, albeit approximate,
description of a non-singular one, the physical meaning of the
observables should survive the limiting
procedure. As we have demonstrated, such a viewpoint is consistent
with localized dynamics in the presence of traps modelled by
impenetrable barriers.

There is one common feature shared by the models considered in
Secs.~III--V: the Hamiltonian is a well defined self-adjoint
operator in each case, respecting various confinement requirements
by suitable boundary conditions. There is however no consistent
canonical quantization procedure that can be carried out
exclusively in the trap interior, because in the case of Dirichlet
boundary conditions there is no self-adjoint momentum-like
operator. If we do not ignore the exterior of the trap the
momentum observable paradoxes disappear and the canonical
quantization procedure reduces to its textbook meaning also in the
presence of impenetrable barriers.

\begin{acknowledgments}
We would like to thank J. Piskorski and H. Falomir for comments.
This research has been supported by the Polish Ministry of Scientific
Research and Information Technology under the grant No
PBZ-MIN-008/P03/2003.
\end{acknowledgments}

\appendix
\section{Basic mathematical concepts}

We shall give a brief introduction to the basic mathematical concepts
employed in the paper, with an emphasis on the distinctions
between symmetric and self-adjoint operators in Hilbert space.

(1) {\sl Absolute continuity}. Let $\phi(x)$ be locally integrable
on $R$. Then $f(x) = \int_a^x \phi (t) dt$ is called absolutely
continuous and denoted by $f \in AC(R)$. If $\phi $ is continuous,
then $f$ is differentiable and ${\frac{d f(x)}{dx}} = \phi (x)$.
If ${\frac{d}{dx}}$ is understood as an operator in Hilbert space
and its domain contains absolutely continuous functions, then we
set ${\frac{d f(x)}{dx}} = \phi (x)$, even if $f$ happens not to
be differentiable.

(2) {\sl Domains of operators}. Most of the operators discussed
in this paper are unbounded. When defining an unbounded operator,
it always is necessary to specify its domain of definition. If $A$
is an operator in the Hilbert space ${\cal{H}}$, we write $D(A)
\subset {\cal{H}}$ for the domain of $A$. An operator $B$ is
called an extension of $A$, which is often written as $A \subset
B$, if and only if $D(A) \subset D(B)$ and $Af = Bf$ for all $f\in
D(A)$.

(3) {\sl Symmetric versus self-adjoint operators}. An operator $B$
is adjoint to $A$ if (a) $(g,Af) = (Bg,f)$ for all $f\in D(A)$ and
$g\in D(B)$, (b) $B$ is a maximal operator with the property (a),
in the sense that if $B\subset C$ and $B\neq C$, then (a) does not
hold for $C$. We write $B= A^*$ if $B$ is adjoint to $A$. It
follows that $A\subset C$ implies $C^* \subset A^*$. We say that
$A$ is symmetric if $A \subset A^*$ and self-adjoint if $A=A^*$.

(4) {\sl Closed operator}. Let us consider a densely defined
operator $A$. For any $g\in D(A)$, we set $||g||_1 = [(Ag,Ag) +
(g,g)]^{1/2}$. Then $||\cdot ||_1$ is a norm in $D(A)$. If $f_n
\in D(A)$ is a Cauchy sequence in $||\cdot ||_1$, that is,
$\lim_{n,m \rightarrow \infty} ||f_m - f_n||_1 =0$, then $f_n$
also is a Cauchy sequence in the Hilbert space ${\cal{H}}$ norm
$||f|| = [(f,f)]^{1/2}$. By the completeness of ${\cal{H}}$ there
is $f\in {\cal{H}}$ such that $\lim_{n\rightarrow \infty}
||f-f_n|| = 0$. If it follows that $f$ necessarily belongs to
$D(A)$ (that is, $D(A)$ is complete in the $||\cdot ||_1$ norm),
then we say that $A$ is closed and we write $A = \overline{A}$. If
$A$ is not closed, it still may have a closed extension. That can
be guaranteed by assuming $D(A^*)$ to be dense in ${\cal{H}}$.

Under such circumstances the $||\cdot||_1$-norm limit
$\lim_{n\rightarrow \infty} Af_n$ exists for any Cauchy sequence
$f_n\in D(A)$ and moreover $g= \lim_{n\rightarrow \infty} Af_n$ is
the same for all sequences $f_n$ converging to the same limit $f$.
Thus we may define $\overline{A}f = \lim_{n\rightarrow \infty}
Af_n$ . The operator $\overline{A}$ is a minimal closed extension
of $A$; $\overline{A}$ is called a closure of $A$. We have
$A^*=(\overline{A})^*$, $\overline{A^*}=A^*$.

(5) {\sl Self-adjoint extension}. Let $A$ be symmetric, $A
\subset A^*$ but is not necessarily self-adjoint. The closure
$\overline{A}$ of $A$ obeys $A\subset \overline{A}\subset A^*$.
Even if $A\neq A^*$, we may have $\overline{A} = A^*$ and then
$A$ is called essentially self-adjoint. However, typically we may
expect that $A^* \neq \overline{A}$ and at this point we need to
invoke the notion of the self-adjoint extension.

Suppose that $B$ is a symmetric extension of $\overline{A}$, then
$\overline{A} \subset B\subset B^* \subset A^*$. Can we extend
$\overline{A}$ so that $\overline{A} \subset B=B^* \subset A^*$,
that is, has $\overline{A}$ a self-adjoint extension? If so, is
this extension unique? The full answer to those questions is given
by the Krein-von Neumannn theory of self-adjoint extensions of
symmetric operators\cite{glazman} which we shall invoke in the following.

(6) {\sl Deficiency indices and self-adjoint extensions}. Let $A$
be a closed operator, that is, $A=\overline{A}$. We denote by
${\cal{M}}, {\cal{N}} \subset {\cal{H}}$ the spaces of the
solutions of $(A^* \mp i)g = 0$ and by $m$ and $n$ respective
dimensions of these spaces. The numbers $n,m$ are called
deficiency indices for $A$. For simplicity, we assume $m$ and $n$ to be
finite. Then, $A$ has self-adjoint extensions if and only if
$n=m$. Let the deficiency indices of $A$ form a pair $(n,n)$. Then
there is a one-to-one correspondence between the self-adjoint
extensions of $A$ and the family of all unitary $n \times n$
matrices. We consider some examples in the following.

(a) Consider ${\cal{H}} = L^2(a,b)$ and $A=
-i{\frac{d}{dx}}$ acting in
$D(A) = C_0^{\infty}(a,b) \subset L^2(a,b)$. We recall that $f\in
C_0^{\infty}(a,b)$ if and only if $f$ is infinitely differentiable and
${\rm supp}\, f \subset (a,b)$. Accordingly, $\overline{A}
=-i{\frac{d}{dx}}$ with the domain $D(\overline{A}) =
\{ f\in AC(a,b); f(a) = f(b) =0\}$. Integration by parts shows that
$A^* =
\overline{A}^* =
-i{\frac{d}{dx}}$ with $D(A^*) = AC(a,b)$. Thus $\overline{A} \subset
A^*$, that is,
$\overline{A}$ is a closed symmetric operator and the equations $(A^*
\mp i)g = 0$ take
the form $(-i{\frac{d}{dx}} \mp i)g = 0$. The solutions are $\exp (\mp
x)$, and hence $m= \dim {\cal{M}} =
\dim {\cal{N}} = 1$, and the family of self-adjoint extensions is
indexed by $\exp (i\alpha)$ with $0\leq \alpha <2 \pi$. The
self-adjoint extensions are determined in terms of the boundary
conditions; $A_{\alpha} = A^*_{\alpha} = -i{\frac{d}{dx}}$ with
respective domains $D(A_{\alpha}) = \{ f \in AC(a,b); \, f(a) =
\exp (i\alpha) f(b)\}$.

(b) Consider $H = -{\frac{d^2}{dx^2}}$ with $D(H) =
C_0^{\infty}(a,b)$. Then we have $\overline{H}
=-{\frac{d^2}{dx^2}}$ with the domain $D(\overline{H}) =\{ f\in
AC^2(a,b); \, f(a) = f(b) = f'(a) = f'(b)=0 \}$, where $AC^2(a,b)$
denotes functions with absolutely continuous first derivatives. Two
integrations by parts show that $\overline{H}$ is symmetric and
$H^* =-{\frac{d^2}{dx^2}}$ acts in the domain $D(H^*) =
AC^2(a,b)$. The deficiency indices of $H^*$ follow from
$(-{\frac{d^2}{dx^2}} \mp i)g = 0$. In both cases we obtain the same
pair of linearly independent solutions: $\exp(\pm kx)$ with
$k=(1-\sqrt{2})(1+i)/\sqrt{2}$. Therefore, ${\cal{M}} = {\cal{N}}$
and $m=n=2$.

(c) Now let $a=0$ and $b=\infty$, that is, ${\cal{H}} =
L^2(0,\infty)$. In this case, $\exp(x)$ is not an element of
${\cal{H}}$, and $\exp (-x) \in {\cal{H}}$. Thus $m=0$ and $n=1$,
and hence there is no self-adjoint extension of $A =
-i{\frac{d}{dx}}$. On the other hand, the same reasoning for
$\overline{H}$ implies that $m=n=1$, and thus there is a
one-parameter family of self-adjoint extensions on the half-line.

(d) If we choose $a= -\infty$ and $b= +\infty $, that is,
${\cal{H}} = L^2(R)$, we have $m=n=0$ for both $\overline{A}$ and
$\overline{H}$. Therefore in this case, both $A$ and $H$ are
essentially self-adjoint.

(7) {\sl Spectral theorem}. The spectral theorem describes
self-adjoint operators in terms of projection operators. We shall
describe how it works for operators discussed in the paper.

For each $0\leq \alpha < 2\pi $ the family $\{ e^{\alpha}_n(x);
n=0,\pm1,\pm2, \ldots\}$ defined by Eq.~(\ref{three}) is an
orthonormal basis in $L^2([0,\pi])$. We denote by $Q_n^{\alpha}$ the
projection operator onto the one dimensional space spanned by the
$e^{\alpha}_n(x)$. The operator $P_{\alpha}$, Eq.~({\ref{two}), can be written
as $P_{\alpha} = \sum_{n=-\infty}^{n=+\infty} (2n+ {\frac{\alpha}{\pi}})
Q_n^{\alpha}$. The condition for $f$ to be in the domain $D(P_{\alpha})$ of
$P_{\alpha}$ follows by direct calculation, see for example, our comment
below Eq.~(\ref{four}). Now, we can define functions of
$P_{\alpha}$, for example $H_{\alpha}= P^2_{\alpha} =
\sum_{-\infty}^{+\infty} (2n+ {\frac{\alpha}{\pi}})^2
Q^{\alpha}_n$ with $D(P_{\alpha}^2)$ given by Eq.~(\ref{domain}).
Similarly $\exp(-iH_{\alpha}t) = \sum_{-\infty}^{+\infty}\exp[-i(2n+
{\frac{\alpha}{\pi}})^2t] \, Q_n^{\alpha}$. Note that although both
$P_{\alpha}$ and $P^2_{\alpha}$ are unbounded, the operator
$\exp(-iP_{\alpha}^2t)$ is bounded and defined on the whole of
$L^2([0,\pi])$.

(8) {\sl Momentum representation}. We introduce the notion
$\tilde{P}$ of the ``momentum operator in the momentum
representation": $\tilde{P}f(p) =pf(p);\, D(\tilde{P})=\{ f\in
L^2(R); \int |pf(p)|^2dp <\infty \}$. We also have
$\tilde{P}^2f(p) =p^2f(p);\, \, D(\tilde{P}^2)=\{ f\in L^2(R);
\int |p^2f(p)|^2dp <\infty \}$. The operator
$\exp(-i\tilde{P}^2t)f(p) = \exp(-ip^2t)f(p)$ is bounded and
defined on the whole of $L^2(R)$.

If ${\cal{F}}$ stands for the Fourier transformation and
${\cal{F}}^{-1}$ for its inverse, then $P={\cal{F}}^{-1} \tilde{P}
{\cal{F}}$ and $D(P) = {\cal{F}}^{-1}D(\tilde{P})$. Analogously we
have $P^2={\cal{F}}^{-1}\tilde{P}^2{\cal{F}};\, \, D(P^2) =
{\cal{F}}^{-1}D(\tilde{P}^2)$ and $\exp(-iP^2t)=
{\cal{F}}^{-1}\exp(-i\tilde{P}^2t){\cal{F}}$.

\end{document}